\documentclass[review]{elsarticle}

\usepackage{lineno,hyperref,epsfig}
\modulolinenumbers[5]

\journal{Journal of Magnetism and Magnetic Materials}

\begin{document}

\begin{frontmatter}

\title{Orbital-selective behavior in Y$_5$Mo$_2$O$_{12}$ and (Cd,Zn)V$_2$O$_4$}

\author{Sergey V. Streltsov}
\address{Institute of Metal Physics, S.Kovalevskoy St. 18, 620990 Ekaterinburg, Russia}
\address{Ural Federal University, Mira St. 19, 620002 Ekaterinburg, Russia}

\begin{abstract}
We present two examples of the real materials, which show orbital-selective 
behavior. In both compounds a part of the electrons is localized
on the molecular orbitals, which lead to a significant reduction of 
the magnetic moment on the transition metal ion.
\end{abstract}


\end{frontmatter}

\linenumbers

\section{Introduction}
It was recently shown that there may exist so called orbital-selective state
in the dimerized systems with orbital degrees of freedom~\cite{Streltsov2014}.
This is the state in which different orbitals behave in different manners.
All electrons (orbitals) in the orbital-selective state are split
on two qualitatively different groups. One part of the electrons occupy
molecular orbitals and form spin singlets being magnetically 
inactive, while other electrons are effectively decoupled from them and
have local magnetic moments (which can be ordered, i.e. lead
to formation of ferro- or antiferromagnetic states, or disordered, i.e.
paramagnetic). As a result one of the main features of 
the orbital-selective state is substantially reduced 
magnetic moment. Because of the formation of molecular orbitals this moment turns
out to be much smaller, than expected basing on the
purely ionic model consideration.

In the present paper we show that the orbital-selective state
is realized in two real materials: Y$_5$Mo$_2$O$_{12}$ and ZnV$_2$O$_4$ (and 
also in CdV$_2$O$_4$, which is iso-electronic and iso-structural to 
ZnV$_2$O$_4$).

\section{Calculation details}
We used the density functional theory within the generalized 
gradient approximation (GGA)~\cite{Perdew1996} to study electronic properties of Y$_5$Mo$_2$O$_{12}$.
We used full-potential WIEN2k code~\cite{Blaha2001}.
The radii of atomic spheres were set as following $R_{Y}=2.19$~a.u., $R_{Mo}=1.88$~a.u.,
and $R_{O}=1.70$~a.u. The Brillouin-zone (BZ) integration in the course of the
self-consistency was performed over a mesh of
200 {\bf k}-points. The parameter of the plane wave expansion
was chosen to be $R_{MT}K_{max} = 7$, where $R_{MT}$ is the
smallest atomic sphere radii and $K_{max}$ - plane wave
cut-off.

\section{Y$_5$Mo$_2$O$_{12}$}
\begin{figure}[b!]
\begin{center}
 \includegraphics[clip=false,width=0.3\textwidth]{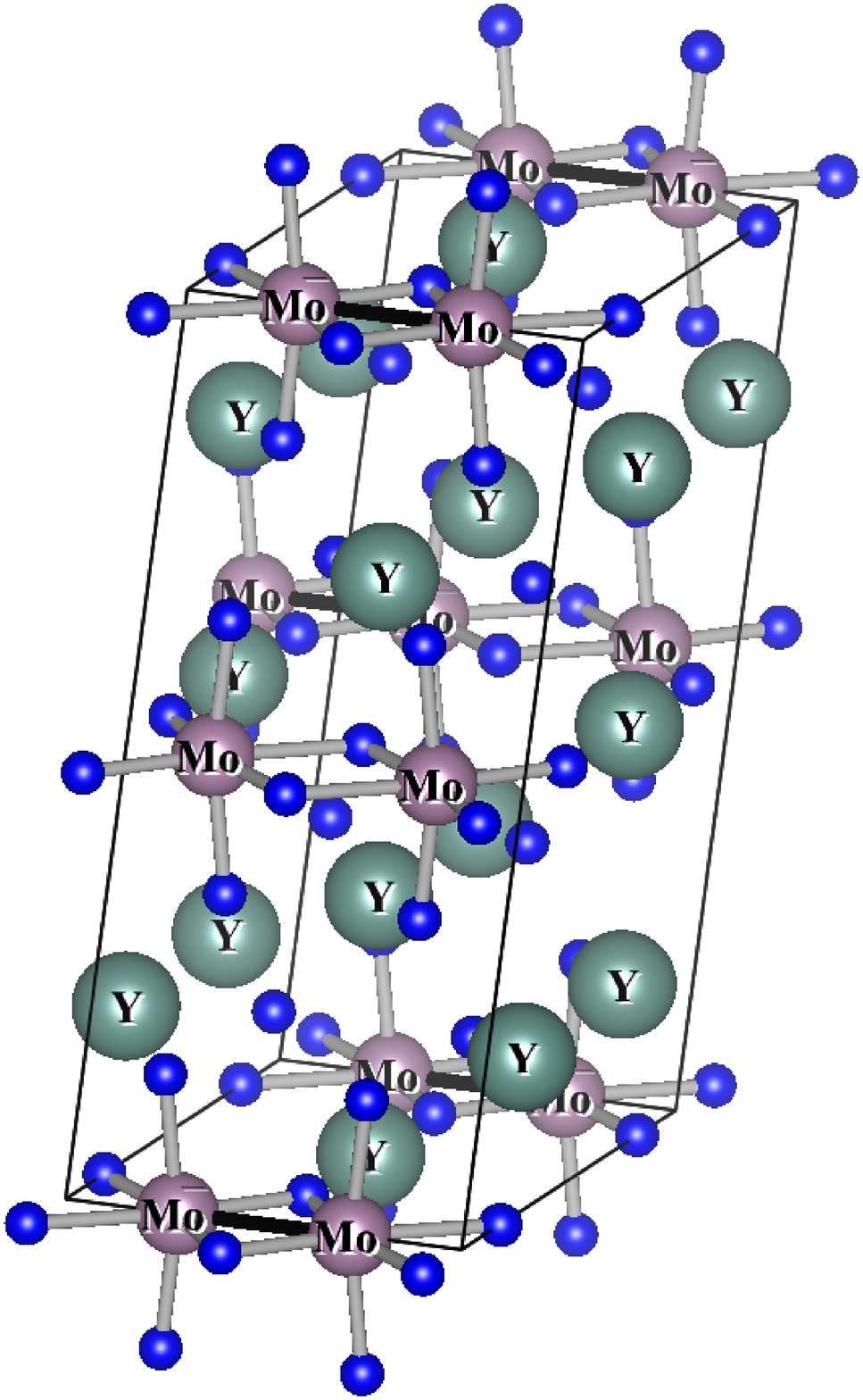}
 \includegraphics[clip=false,width=0.5\textwidth]{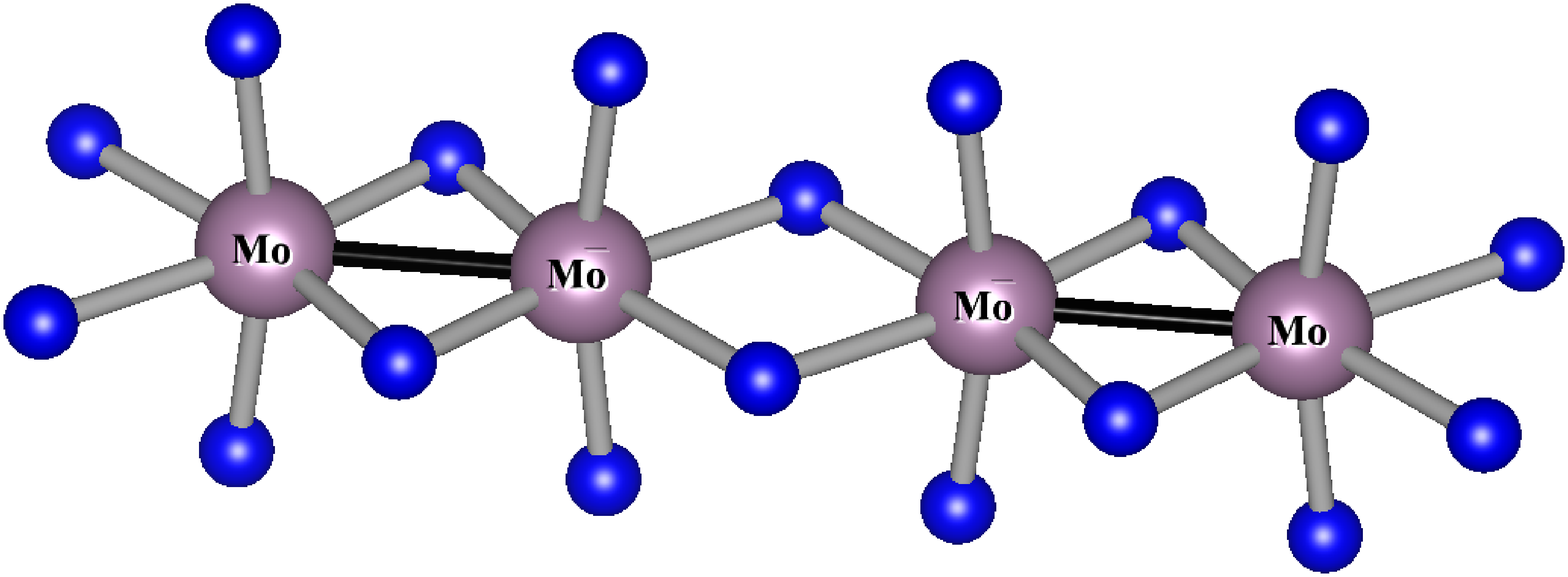}
\end{center}
\caption{\label{YMoO-str} Left: the unit cell of Y$_5$Mo$_2$O$_{12}$; Mo 
ions are shown in violet, O $-$ in blue, and Y $-$ in grayish green.
Right: main structural element of Y$_5$Mo$_2$O$_{12}$ $-$ dimerized Mo
chain.}
\end{figure}

The crystal structure of Y$_5$Mo$_2$O$_{12}$ is formed by the edge sharing 
MoO$_6$ chains, which are dimerized. These chains are separated by the Y ions, see 
Fig.~\ref{YMoO-str}. The electronic configuration of Mo having formal valence 
4.5+ is $4d^{1.5}$. Hence one may expect that the local magnetic moment
would be $\mu_{loc}=1.5\mu_B$/Mo, while effective magnetic moment
in the Curie-Weiss theory will be $\mu_{eff}=2.29\mu_B$/Mo or 
$\mu_{eff}=1.15\mu_B$/dimer. Experimentally, however,
effective moment equals $\mu_{eff}=1.7\mu_B$/Mo~\cite{Torardi1985}.
We attribute this feature to the fact that this compound is
actually in the orbital-selective state.

\begin{figure}[t!]
\begin{center}
 \includegraphics[clip=false,angle=270,width=0.7\textwidth]{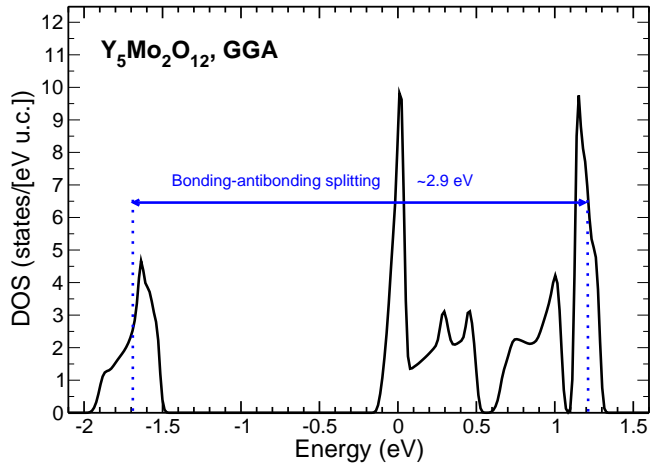}
\end{center}
\caption{\label{YMoO-DOS} Total density of states of the Y$_5$Mo$_2$O$_{12}$,
obtained in the nonmagnetic GGA calculation.}
\end{figure}

The total density of states obtained in the nonmagnetic GGA calculation is 
shown in Fig.~\ref{YMoO-DOS}. In the edge-shared geometry there are two
types of the $t_{2g}$ orbitals: $\sigma$-bonded $xy$ orbitals
(which look directly to each other) and $\pi$-bonded $yz$ and 
$zx$ orbitals on different Mo centers (it is assumed that the local $z$ axis
is perpendicular to the plane defined by the Mo-Mo bond and common edge for two
MoO$_6$ octahedra). Direct overlap between the $xy$ orbitals on two Mo sites 
leads to huge bonding-antibonding splitting $\sim$2.7 eV. Therefore one may
gain energy occupying bonding orbital with two electrons (total number of
electron is 3 per dimer). Remaining electron will provide net magnetic moment of
the dimer, giving $S=1/2$ per dimer. This yields $\mu_{eff}=1.74\mu_B$/Mo,
which agrees with experimental data~\cite{Torardi1985}.
Resulting state is orbital-selective and is realized due to rather 
uncommon set of the parameters: one of the hopping parameter
$t_{xy/xy} \sim 1.35$~eV is much larger both than intra-atomic Hund's
rule exchange (which is estimated for early $4d$ transition metals
to be $\sim$0.7 eV~\cite{Lee2006,Streltsov2012a}) and
hopping integrals between other orbitals $(t_{xz/xz},t_{yz/yz})\ll t_{xy,xy}$. 

%%%%%%%%%%%%%%%%%%%%%%%%%%%%%%%%%%%%%%%%%%%%%%%%%%%%%%%%%%%%%%%%%%%%%%%%
\section{CdV$_2$O$_4$ and ZnV$_2$O$_4$}
The electronic and magnetic properties of the V spinels are thoroughly
investigated last years.~\cite{Giovannetti2011,Pardo2008} E.g. it was found that CdV$_2$O$_4$ 
is multiferroic below $T_N=$33~K~\cite{Giovannetti2011}. The mechanism of the
multiferroicity is the magnetostriction: unconventional 
$\uparrow \uparrow \downarrow \downarrow \uparrow \uparrow$ magnetic
order results in the dimerization of the V ions having the same spin projection (i.e. 
$\uparrow \uparrow$). This in turn leads to the shift of the oxygen ions away 
from the high symmetry positions and onset of the spontaneous 
electric polarization~\cite{Giovannetti2011}. The missing element in this 
microscopic model is why the spins of V ions forming dimers show ferromagnetic
order. We know from the basics of the quantum mechanics that
the spin singlet (i.e. antiferro ordering of the spins, as e.g. in
the hydrogen molecule) is more typical for dimers.

We argue that it can be related with the orbital-selective physics.
The situation here is rather similar to Y$_5$Mo$_2$O$_{12}$, since
neighboring VO$_6$ octahedra also share their edges. There is again strong 
overlap between the $xy$ orbitals. Since V$^{3+}$ has $3d^2$ electronic
configuration, there are four $d$ electrons per dimer. Two
electrons may occupy molecular orbital of the $xy$ symmetry,
while other two stay on the $yz$ and $zx$ local orbitals and
provide $S=1$ per V dimer, or $S=1/2$ per V. Corresponding
exchange constant (added in the revised version of the manuscript) in this short 
V-V pair in CdV$_2$O$_4$ is ferromagnetic and equals $J=56$~K as calculated in the LSDA+U
method for $U-J_H$=0.7~eV using the Green's function 
formalism~\cite{Korotin2014,LEIP1}.

The distortions of the crystal structure 
and type of the magnetic structure in ZnV$_2$O$_4$ are the same as in 
CdV$_2$O$_4$~\cite{Pardo2008}, but there is an additional information
about magnetic properties of ZnV$_2$O$_4$. The local magnetic
moment on V was found to be 0.66~$\mu_B$ in ZnV$_2$O$_4$~\cite{Reehuis2003}.
This qualitatively agrees with the model treatment presented above. In the
ionic model this corresponds to the local spin moment $\mu = gS = 1\mu_B$,
but this value is further reduced due to hybridization 
effects, which are always exist in real materials.
\begin{figure}[t!]
\begin{center}
 \includegraphics[clip=false,width=0.7\textwidth]{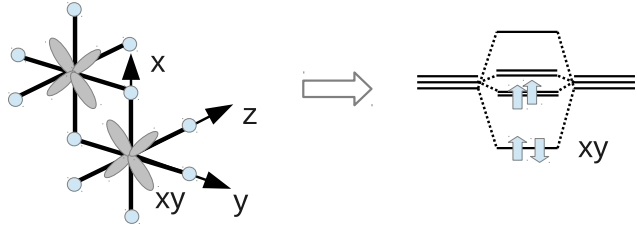}
\end{center}
\caption{\label{CdV2O4}Left: the pair of the $xy$ orbitals in V dimer, which
have the largest hopping parameters in the system. Right: sketch illustrating
energy levels splitting in (Cd,Zn)V$_2$O$_4$.}
\end{figure}

\section{Conclusions}
In the present paper we show two examples of the
orbital-selective behavior, which results in the suppression
of the magnetic moment in (Cd,Zn)V$_2$O$_4$ and Y$_5$Mo$_2$O$_{12}$.

\section{Acknowledgments}
Author is grateful to D. Khomskii with whom the investigation of the
orbital-selective state was performed. The calculation of exchange constants 
is supported by the Russian Science Foundation 
via  RNF 14-22-00004. 

\section*{References}

%\bibliography{mybibfile}
\bibliography{../library}
\end{document}